\documentclass[twocolumn,showpacs,prl,amsmath,amssymb]{revtex4} 
\usepackage{graphicx}
\usepackage{dcolumn}  
\usepackage{bm}       
\usepackage{psfig,pstricks,pst-grad,fancybox,graphics}
\usepackage{latexsym}
\usepackage{amsmath}
\usepackage{amssymb}
\usepackage{enumerate}
\usepackage{bbm}
\usepackage[latin1]{inputenc}

\newcommand{\R}{\ensuremath{\mathbbm R}}

\newcommand{\ket}[1]{\ensuremath{|#1\rangle}}

\newcommand{\braket}[2]{\ensuremath{\langle #1|#2\rangle}}

\newcommand{\ketbra}[1]{\ensuremath{| #1 \rangle \langle #1 |}}

\newcommand{\Eins}{\ensuremath{\mathbbm 1}}
\newcommand{\eins}{\ensuremath{\mathbbm 1}}

\newcommand{\BE}{\begin{equation}}
\newcommand{\EE}{\end{equation}}
\newcommand{\kommentar}[1]{}

\newcommand{\mean}[1]{\ensuremath{\langle #1 \rangle}}
\newcommand{\qed}{\ensuremath{\hfill \Box}}

\begin{document}

\title{Characterizing Entanglement via Uncertainty Relations }
\author{Otfried Gühne}
\affiliation{Institut für Theoretische Physik, Universität Hannover,
Appelstraße 2, D-30167 Hannover, Germany}
\date{\today} 
\begin{abstract}
We derive a family of necessary separability criteria for 
finite-dimensional systems based on inequalities for variances of 
observables. We show that every pure bipartite entangled state violates 
some of these inequalities. Furthermore, a family of bound entangled states
and true multipartite entangled states can be detected. 
The inequalities also allow to distinguish 
between different classes of true tripartite entanglement for  
qubits. We formulate an equivalent criterion in terms of covariance 
matrices. This allows us to apply criteria known from the regime 
of continuous variables to finite-dimensional systems.
\end{abstract}
\maketitle

The detection of entanglement  is one of the fundamental 
problems in quantum information theory. From a theoretical point 
of view one can try to answer the question whether a given 
entirely known state is entangled or not, but despite a lot of 
progress in the last years \cite{criteria}, no general
solution of this problem is known. In experiments, one aims 
at detecting entanglement without knowing 
the state completely. Bell inequalities \cite{bell} and entanglement 
witnesses \cite{witnesses} are the main tools to tackle
this task. Both rely on inequalities for {\it mean values} 
of observables, and violation of them implies entanglement. 
Entanglement criteria based on inequalities for {\it variances} of 
observables have also been studied, mainly designed for continuous 
variable systems \cite{alle}. Recently, it has been shown 
that variances of observables of a special type can detect 
entanglement also in finite dimensional bipartite systems \cite{hofmann1}. 
Even an example of bound entanglement, allowing detection in 
this way, has been found \cite{hofmann2}.  

In this paper, we  present a generalization of this 
approach by considering arbitrary observables. This leads to new 
results in various directions. First, we show how the entanglement 
of every pure bipartite entangled state can be detected with our 
method. Second, we present inequalities for the detection of a class 
of bound entangled states. Third, we show how 
multipartite entanglement can be detected. Our approach even
admits to distinguish different classes of true multipartite 
entanglement. 
Finally, we show that our criteria are equivalent to a criterion in 
terms of covariance matrices. This formulation allows us to translate 
separability criteria  known from continuous variables
\cite{werner, giedke1} to finite-dimensional systems. 

We start with the following observation. Let $\varrho$ be 
a density matrix, and let $M$ be an observable. We denote 
the expectation value of $M$ by $\mean{M}_{\varrho} := Tr(\varrho M)$
and the variance (or uncertainty) of $M$ by 
\begin{equation}
\delta^2(M)_{\varrho} := \mean{(M-\mean{M}_{\varrho})^2}_{\varrho}
= \mean{M^2}_{\varrho}-\mean{M}^2_{\varrho}.
\end{equation}
We suppress the dependence on $\varrho$ in our notation, when 
there is no risk of confusion. If $\varrho$ is a pure state 
the variance is zero iff $\varrho$ is an eigenstate
of $M.$ Now we have:
 
{\bf Lemma 1.} Let $M_i$ be some observables and 
$\varrho=\sum_k p_k \varrho_k$ be a convex combination 
({\it i.e.} $p_k\geq 0, \sum_k p_k =1$) of some states 
$\varrho_k$ within some subset $S.$ Then
\begin{equation}
\sum_i \delta^2(M_i)_{\varrho} 
\geq 
\sum_k p_k \sum_i \delta^2(M_i)_{\varrho_k}
\label{l1a}
\end{equation}  
holds. We call a state ``violating Lemma 1'' iff there 
are no states $\varrho_k \in S$ and no $p_k$ such that
Eq. (\ref{l1a}) is fulfilled.

\emph{Proof.} This fact is known, see e.g.  
\cite{galindo, hofmann1}. The inequality holds for each $M_i$: 
$\delta^2(M_i)_{\varrho} = 
\sum_k p_k \mean{(M_i-\mean{M_i}_{\varrho})^2}_{\varrho_k}
=\sum_k p_k 
(\delta^2(M_i)_{\varrho_k}
+
(
\mean{M_i}_{\varrho_k}
-
\mean{M_i}_{\varrho}
)^2
)
\geq \sum_k p_k \delta^2(M_i)_{\varrho_k}.$
$\qed$ 

Please note that this Lemma has a clear physical meaning: One cannot
decrease the uncertainty of an observable by mixing several states. 

In most cases we will be interested in separable states. They can
be written as a convex combination of product states. So, unless 
stated otherwise, we will assume that in Lemma 1 the set  $S$ 
denotes the set of 
product states. Then violation of this Lemma implies 
entanglement of the state $\varrho,$ which can be detected with uncertainties. 
In fact, the main idea of this paper is to show that Eq. 
(\ref{l1a}) yields strong sufficient entanglement criteria, 
if we choose the $M_i$ appropriately. 

For completeness, let us remind the reader of the so-called 
``Local Uncertainty Relations'' (LURs), introduced by Hofmann 
and Takeuchi \cite{hofmann1}. Let $A_i$ be observables 
on Alice's space  of a bipartite system. If they do not share 
a common eigenstate, there is a number $C_A > 0$ such that 
$\sum_i \delta^2(A_i)_{\varrho_A}\geq C_A$ holds for all 
states $\varrho_A$ on Alice's space. Hofmann and Takeuchi showed:  

{\bf Proposition 1.} Let $\varrho$ be separable and let
$A_i,B_i,  i=1,...,n$  be operators on 
Alice's (resp. Bob's) space, fulfilling 
$\sum_{i=1}^n \delta^2(A_i)_{\varrho_A}\geq C_A$ and 
$\sum_{i=1}^n \delta^2(B_i)_{\varrho_B}\geq C_B.$ We define
$M_i := A_i \otimes \Eins +\Eins \otimes B_i.$ 
Then 
\begin{equation}
\sum_{i=1}^n\delta^2(M_i)_{\varrho} \geq C_A + C_B
\end{equation}
holds.

The LURs provide strong criteria  which can by 
construction be implemented with local measurements.
Nevertheless, they have some disadvantages: it is not 
clear which operators $A_i$ and $B_i$ one should 
choose to detect a given entangled state. 
Also, LURs can by construction characterize separable states only; 
they do not apply for other convex sets. 
Finally, it is not clear 
how to generalize them to multipartite systems. In fact, 
no LUR for the detection of true multipartite 
entanglement in finite dimensional systems 
is known so far.

A way to overcome these disadvantages is to consider  
\emph{nonlocal} observables. For an experimental implementation
one can decompose any nonlocal observable into local operators {\it i.e.} 
write it as a sum of projectors onto product vectors \cite{exdet}. Each 
of the terms in this sum can then be measured locally. 
The measurement of a nonlocal observable in this way
has recently been implemented \cite{fdm}.
Let us start with the case of two qubits: 

{\bf Proposition 2.} Let $\ket{\psi_1}=a\ket{00}+b\ket{11}$ be an 
entangled two qubit state written in the Schmidt decomposition, with
$a\geq b.$ 
Then, there exist $M_i$ such that for
$\ket{\psi_1}$, $\sum_i \delta^2(M_i)_{\ketbra{\psi_1}} = 0$
holds, while for separable states
\begin{equation}
\sum_i \delta^2(M_i) \geq 2 a^2 b^2
\label{p2a}
\end{equation}
is fulfilled. 
The $M_i$ are explicitly given (see below). 

\emph{Proof.} We define
$ \ket{\psi_2} =  a\ket{01}+b\ket{10}
; \;\;
\ket{\psi_3} =  b\ket{01}-a\ket{10}
; \;\;
\ket{\psi_4} =  b\ket{00}-a\ket{11},$ and further 
$M_i := \ketbra{\psi_i}, i=1,...,4.$ Then  
$\sum_i \delta^2(M_i)_{\ketbra{\psi_1}} = 0.$
We only need to prove the bound (\ref{p2a}) for 
a pure product vector $\ket{\phi}.$ We have 
$\sum_{i=1}^4 \delta^2(M_i)_{\ketbra{\phi}}
= 1 - \sum_i (|\braket{\phi}{\psi_i}|^2)^2.$
Now, we use the fact that $|\braket{\phi}{\psi_i}|^2 \leq a^2.$
To see this, we expand $\ket{\psi_i}$ in a product basis 
$\ket{\psi_i}=\sum_{kl}C^{i}_{kl}\ket{kl}$ and also any 
product vector: $\ket{a,b}=\sum_{kl}a_k b_l \ket{kl}.$ 
Then $\max_{\ket{a,b}}|\braket{a,b}{\psi_i}| 
= \max_{a_m,b_n} |\sum_{m,n}a^*_m C^{i}_{mn}b_n |
= \max\{\mbox{sv}(C^{i})\}$ holds,  where $\mbox{sv}(C^{i})$ 
are the singular values 
of $C^{i},$ {\it i.e.} the Schmidt coefficients of $\ket{\psi_i}.$ 
With this bound we have  
$\sum_i (|\braket{\phi}{\psi_i}|^2)^2 \leq (a^2)^2+(1-a^2)^2$
and finally
$\delta^2(M_i)_{\ketbra{\phi}} \geq 2a^2b^2.$
$\qed$

Note that Eq. (\ref{p2a}) allows to detect a mixed state of 
the form $\varrho(p)=p\ketbra{\psi_1}+(1-p) \eins/4$ for 
$ p \geq \sqrt{1-{8a^2b^2}/{3}}.$

This two-qubit example reveals already a main idea for the 
construction of the $M_i$: We take one $M_i$ as the projector 
onto the range of $\varrho$, and the others as projectors onto a 
basis of the kernel. This construction 
suffices to detect all bipartite pure entangled states, 
many bound entangled states and multipartite 
entanglement. 

{\bf Proposition 3.} Let $\ket{\psi_1}$ be an entangled pure
state in a bipartite $N\times M$-system. Then $\ket{\psi_1}$ 
violates Lemma 1 for properly chosen $M_i.$ 

\emph{Proof.} Let $U$ be the space orthogonal to 
$\ket{\psi_1}.$ It is clear that $U$ contains 
at least one entangled vector $\ket{\psi_e}.$ We can
choose a basis $\ket{\psi_i}, i=2,...,NM$ of $U$ 
which consists only of entangled vectors.  
(To do this, we choose an arbitrary, not necessarily orthogonal, 
basis. If it contains product vectors, we perturb them by adding  
$\varepsilon \ket{\psi_e} $). Then we take
$M_i=\ketbra{\psi_i}.$ The only possible common eigenstates 
of the $M_i$ are the $\ket{\psi_i}.$ So for product states
the sum over all uncertainties is bounded from below, while
it is zero for $\ket{\psi_1}.$
$\qed$

It is interesting that the existence of an entangled basis 
in the kernel of a state $\varrho$ suffices to derive 
uncertainty relations also for a class of bound entangled 
states.
Here, the class of bound entangled states which we 
want to consider are those arising from an unextendible
product basis (UPB) \cite{upb1}. They can be constructed 
as follows: Let $\{ \ket{\phi_i}=\ket{e_i}\ket{f_i}, i=1,...,n\}$ 
be a UPB, {\it i.e}. all $\ket{\phi_i}$ are (pairwise orthogonal) product 
vectors, not spanning the whole space,
and there is no product vector orthogonal to all 
$\ket{\phi_i}.$ Then the state
\begin{equation}
\varrho_{UPB}:= \mathcal{N}(\eins - \sum_i \ketbra{\phi_i})
\end{equation}
is a bound entangled state. $\mathcal{N}$ denotes the 
normalization.

{\bf Proposition 4.} Let $\varrho_{UPB}$ be a bound entangled 
state, constructed from a UPB. Then $\varrho_{UPB}$ violates 
Lemma 1 for appropriate
$M_i.$

\emph{Proof.} Let $U$ be the subspace spanned by the UPB. Then
$U$ must contain at least one entangled vector. To see this, 
note that subspaces containing only product vectors are of 
the form $\{\ket{v}= \ket{a}\ket{b}\}$ with a fixed $\ket{a}$
(or $\ket{b}$) for all $\ket{v}.$ But then there would be 
product vectors orthogonal to $U.$ Due to the existence of an 
entangled vector in $U,$ there is, according to the proof of Proposition 3,
an entangled basis $\ket{\psi_i}; i=1,...,n$ of $U.$ We take 
$M_i=\ketbra{\psi_i}; i=1,...,n$ and 
$M_{n+1}=\eins - \sum_i \ketbra{\phi_i}.$ The common eigenstates
of the $M_i$ are not product vectors, and we have 
$\sum_i \delta^2(M_i)_{\varrho_{UPB}}=0.$
$\qed$

Before we demonstrate that our method is also useful for  
multipartite systems, we recall some facts about 
three qubit systems \cite{duer1, acin}. A three-qubit state
is called fully separable iff it can be written as 
a convex combination of triseparable 
pure states, and it is called biseparable (BS) iff it is a convex 
combination of pure states, 
which are separable with respect to a bipartite splitting. 
Otherwise, it is called fully entangled. There are 
two classes of fully entangled states which are not convertible
into each other by stochastical 
local operations and classical communication. 
These classes are called the GHZ-class and the W-class, and 
the W-class forms a convex set inside the GHZ-class \cite{acin}.
The following inequalities are formulated so, that they can be checked  
with local measurements:

{\bf Proposition 5.} Let $\varrho$ be a three qubit state. We define
a sum of variances as 
\begin{eqnarray}
E(\varrho) & := &
1- 1/8 \left(
\mean{ \Eins \otimes \Eins \otimes \Eins}^2+
\mean{ \Eins \otimes \sigma_z \otimes \sigma_z}^2+
\right. 
\nonumber \\
& & 
+\mean{ \sigma_z \otimes \Eins \otimes \sigma_z}^2+
\mean{ \sigma_z \otimes \sigma_z \otimes \Eins}^2+ 
\nonumber \\
& & 
+\mean{\sigma_x \otimes \sigma_x \otimes \sigma_x}^2 +
\mean{\sigma_x \otimes \sigma_y \otimes \sigma_y}^2 +
\nonumber \\
& & 
\left.+
\mean{\sigma_y \otimes \sigma_y \otimes \sigma_x}^2 +
\mean{\sigma_y \otimes \sigma_x \otimes \sigma_y}^2 \right).
\label{p5a}
\end{eqnarray}
If $E < 1/2$, the state $\varrho$ is fully tripartite entangled. 
If $E < 3/8$, the state $\varrho$ belongs even to the GHZ-class.
For the GHZ-state $\ket{GHZ}=1/\sqrt{2}(\ket{000}+\ket{111})$ 
we have $E=0.$

\emph{Proof.} We define 
the eight states
$
\ket{\psi_{1/5}} = (\ket{000}\pm\ket{111})/\sqrt{2}
; \;\;
\ket{\psi_{2/6}} = (\ket{100}\pm\ket{011})/\sqrt{2}
; \;\;
\ket{\psi_{3/7}} = (\ket{010}\pm\ket{101})/\sqrt{2}
; \;\;
\ket{\psi_{4/8}} = (\ket{001}\pm\ket{110})/\sqrt{2}
$
and as usual $M_i = \ketbra{\psi_i},i=1,...,8$ 
and $E=\sum_{i} \delta^2(M_i).$ 
Then the proof  is quite similar to the proof of 
Proposition 2. One only needs the maximal squared overlaps, 
which are known \cite{acin}: 
$\max_{\ket{\phi}\in BS}|\braket{GHZ}{\phi}|^2 = 1/2$,
$\max_{\ket{\phi}\in W} |\braket{GHZ}{\phi}|^2 = 3/4.$
By decomposing the $M_i$ as in \cite{ustron}, one gets  
$E$ in the form of Eq. (\ref{p5a}).
$\qed$

Let us compare this with the witness for this detection 
problem \cite{acin,ustron}. A witness is given by 
$W=1/2 \cdot \eins - \ketbra{GHZ}$; if $Tr(W\varrho)<0$ the state
$\varrho$ is fully entangled, and if  $Tr(W\varrho) < -1/4$ it even 
belongs to the GHZ-class. Assuming states of the type 
$\varrho(p)=p \ketbra{GHZ}+(1-p)\eins/8$ the method based on $E$
detects them for $p\geq \sqrt{3/7}\approx 0.65$ as tripartite 
entangled, and for $p\geq \sqrt{4/7}\approx 0.76$ as GHZ-states. 
The witness detects them for $p \geq 3/7 \approx 0.43$ as tripartite 
entangled and for $p \geq 5/7 \approx 0.71$ as belonging to the 
GHZ-class. While the witness seems to detect more states in the vicinity 
of $\ket{GHZ},$ the variance method has, in contrast to the 
witness, the property that it detects also all orthogonal 
$\ket{\psi_i}.$ The uncertainty relations are thus in this respect 
stronger that the witness. 

The method presented here can be extended to more parties. 
For four qubits one can write down  16 orthogonal 
GHZ-states of the type
$
\ket{\psi_{i}} = (\ket{x^{(1)}_1 x^{(2)}_1 x^{(3)}_1 x^{(4)}_1}\pm
\ket{x^{(1)}_2 x^{(2)}_2 x^{(3)}_2 x^{(4)}_2})/\sqrt{2}
$
with $x^{(k)}_l \in \{0,1\}$ and  $x^{(k)}_1 \neq x^{(k)}_2.$
If we then define $M_i=\ketbra{\psi_i}$ it follows 
as in the proof of Prop. 5. that for all biseparable 
states $\sum_i \delta^2(M_i)\geq 1/2$ holds. The same 
idea can be used to construct uncertainty relations for 
an arbitrary number of qubits, but the number of $M_i$
increases exponentially.

Now we want to put our results in a more general framework. 
Let $\varrho=\sum_k p_k \varrho^A_k \otimes \varrho^B_k$ be a separable 
state and let $M_i,$ for $ i=1,...,n$ be some observables. 
We can define two functionals $Z(\varrho)[X]$ and $W(\varrho)[X]$ with 
$ X = (x_1,...,x_n) \in \R^n $ by:
\begin{equation}
Z(\varrho)[X]:= \mean {e^{\sum_{i=1}^n x_i M_i}}; 
\;\;\;\;
W(\varrho)[X]:= \ln (Z(\varrho)[X]).
\nonumber
\end{equation}
$Z$ is called the generating functional of the moments and $W$ 
generates the cumulants 
\cite{gernot}. 
We have $Z(\varrho)=\sum_k p_k Z(\varrho^A_k \otimes \varrho^B_k),$ and 
due to the concavity of the 
logarithm it follows that
\begin{equation}
W(\varrho)-\sum_k p_k W(\varrho^A_k\otimes \varrho^B_k) \geq 0
\end{equation} 
holds. The  lhs of this equation has a minimum at $X=0.$ 
Thus,  the matrix of the second derivatives with respect to $X$ 
of the lhs is positive semidefinite at $X=0.$ The matrix of the 
second derivatives of $W$ is the, so-called, covariance matrix 
(CM) $\gamma (\varrho,M_i).$ Its entries are: 
\begin{equation}
\gamma(\varrho, M_i)_{kl} := \partial_k \partial_l W[X] \big|_{X=0},
\label{gammadef}
\end{equation}
where we have denoted $\partial_i := \frac{\partial}{\partial x_i}.$
We will mention some properties of the CM 
later. First, we can state:  

{\bf Lemma 2.} Let $\varrho$ be separable, and let 
$M_i$ be observables. Then there exist 
product states $\varrho^A_k \otimes \varrho^B_k$
and $p_k$ 
such that
\begin{equation}
\gamma(\varrho, M_i)
\geq 
\sum_k p_k \gamma(\varrho^A_k \otimes \varrho^B_k, M_i)
\label{l2a}
\end{equation}
holds. We call a state ``violating Lemma 2'' iff there are no product 
states $\varrho^A_k \otimes \varrho^B_k$ and 
convex weights $p_k,$ such that
Eq. (\ref{l2a}) is fulfilled. 

{\bf Proposition 6.} $\gamma$ has the following properties:
\\
(i) The entries are given by:
\begin{equation}
\gamma_{kl} 
=
(\mean{M_k M_l}+\mean{M_l M_k})/2 -\mean{M_k}\mean{M_l}.
\end{equation}
The diagonal entries are the variances: $\gamma_{ii}=\delta^2(M_i).$
\\
(ii) We have for an arbitrary  $(x_1,...,x_n) \in \R^n $
\begin{equation}
\sum_{i,j=1}^n x_i \gamma_{ij} x_j =\delta^2(\sum_{i=1}^n x_i M_i) 
\geq 0.
\label{p6b}
\end{equation}
In particular we have $\gamma \geq 0.$

\emph{Proof.} (i) can be directly calculated from the definition 
with the help of standard formulas for differentiating operators
\cite{wilcox} and the power expansion of the logarithm. 
(ii) can be proven by calculating $\delta^2(\sum_{i=1}^n x_i M_i).$
$\qed$
 
Of course, one could exhibit other properties of $\gamma,$ but these
two properties suffice for our purpose:

{\bf Theorem 1.} A state $\varrho$ violates Lemma 1 with $S$ being the set 
of separable states iff it violates Lemma 2. The observables leading 
to a violation may differ.

\emph{Proof.} A state violating Lemma 1 violates 
also Lemma 2 with the same $M_i,$ since the sum of all 
variances is the trace of $\gamma.$  To show the other direction, 
let us assume that $\gamma$ violates Lemma 2  and look 
at the set of symmetric matrices of the form 
$T:=\{\sum_k p_k \gamma(\varrho^A_k \otimes \varrho^B_k, M_i) + P\}$
where $P\geq 0$ is positive. $T$ is convex and closed. We have 
$\gamma \notin T,$ and due to a corollary of the Hahn-Banach-Theorem
\cite{scharlau} 
there exists a symmetric matrix $W$ and a number $C$ 
such that $Tr(W\gamma)<C$ while $Tr(W\mu)>C$ for all $\mu \in T$ 
\cite{gezaun}. Since $Tr(WP)\geq 0$ for all $P \geq 0,$ 
we have $W \geq 0.$ Now we
use the spectral decomposition and  write 
$W_{ij} = \sum_l \lambda_l a^l_i a^l_j$ with $\lambda_l \geq 0.$
Let us define $N_l=\sqrt{\lambda_l}\sum_i a^l_i M_i.$
Then $Tr(W\gamma)=\sum_l \delta^2(N_l)<C$ according to 
Eq. (\ref{p6b}), while for all convex combinations of product 
states   
$
\sum_k p_k \sum_l \delta^2(N_l)_{\varrho^A_k \otimes \varrho^B_k} 
>  C
$
holds. This is a violation of Lemma 1 for the $N_l.$
$\qed$

So  Lemma 2 is equivalent to  Lemma 1. 
But the advantage of 
Lemma 2 is that it allows us to relate 
the uncertainty relations to entanglement criteria known from 
continuous variables, mainly Gaussian states. Let us note 
some facts about Gaussian states \cite{werner, giedke1}. The nonlocal 
properties of a Gaussian state are completely encoded in a 
real symmetric CM $\tilde{\gamma},$ which is the CM of the 
canonical conjugate observables position and momentum. A Gaussian 
state $\tilde{\gamma}$ in a bipartite system is separable iff 
there are CMs $\tilde{\gamma}^A, \tilde{\gamma}^B$
\pagebreak
for Alice and Bob such that 
$\tilde{\gamma} \geq \tilde{\gamma}^A \oplus\tilde{\gamma}^B$
\cite{werner, oplus}.
In fact, the separability problem for Gaussian
states was solved in
Ref. \cite{giedke1}, where a map $\tilde{\gamma}_1 \mapsto \tilde{\gamma}_2$ 
was defined which maps a separable (resp. entangled) CM
to another separable (resp. entangled) CM. By iterating 
this map, one gets a CM for which separability is easy to 
check. We can now formulate criteria similar to \cite{werner, giedke1}: 

{\bf Proposition 7.} Let $\varrho$ be separable and let 
$A_i, i=1,...,n$ and $B_i, i= 1,...,m$ be observables on Alice's 
resp. Bob's space. Define $M_i=A_i \otimes \eins, i = 1,...,n$ and 
$M_i=\eins \otimes B_i, i= n+1,...,n+m.$
Then there are states $\varrho^A_k$ and $\varrho^B_k$ and 
convex weights $p_k$ such that if we define 
$\kappa_A := \sum_k p_k \gamma (\varrho^A_k, A_i)$ and 
$\kappa_B  := \sum_k p_k \gamma (\varrho^B_k, B_i)$
the  inequality  
\begin{equation}
\gamma(\varrho, M_i) \geq \kappa_A \oplus \kappa_B
\label{p7a}
\end{equation}
holds. Moreover: We write $\gamma(\varrho, M_i)$ in a block structure:
\begin{equation}
\gamma(\varrho, M_i)=
\left( 
\begin{array}{cc}
A   & C \\
C^T & B
\end{array}
\right),
\label{p8z}
\end{equation}
where $A=\gamma(\varrho, A_i)$ (resp. $B=\gamma(\varrho, B_i)$) 
is an $n \times n$- (resp. $m \times m$)-matrix. 
Then we have \cite{pseudoinverse}:
\begin{eqnarray}
A-C B^{-1} C^T & \geq & \kappa_A
\label{p8a}
\\
B-C^T A^{-1}C & \geq & \kappa_B.
\label{p8b}
\end{eqnarray}

\emph{Proof.} Eq. (\ref{p7a}) follows  from Lemma 2 and the fact
that for these special  $M_i$  for product states: 
$\gamma(\varrho^A \otimes \varrho^B,M_i) = 
\gamma (\varrho^A, A_i) \oplus \gamma (\varrho^B, B_i)$ 
holds.

Eqns. (\ref{p8a}, \ref{p8b}) arise from the first step of the algorithm in
\cite{giedke1}. We arrived already after the first step at a 
block diagonal matrix and thus at a fixed point. 
We prove it as in \cite{giedke1} with a Lemma proven there:
Let $\gamma$ be matrix with a block structure as in (\ref{p8z}). 
Equivalent are \cite{pseudoinverse}: 
(a) $\gamma \geq 0$; 
(b) $ \mbox{ker}(B) \subseteq \mbox{ker}(C)$ 
and $A-CB^{-1}C^T \geq 0$;
(c) $ \mbox{ker}(A) \subseteq \mbox{ker}(C^T)$ 
and $B-C^TA^{-1}C \geq 0$.
Applying the equivalence (a)-(b) to  Eqs. (\ref{p7a}, \ref{p8z}) yields
with Proposition 6, $A-C(B-\kappa_B)^{-1}C^T \geq \kappa_A \geq 0.$ With 
the equivalence (b)-(c) we get $B-\kappa_B-C^TA^{-1}C \geq 0,$ which proves
Eq. (\ref{p8b}). The proof of (\ref{p8a}) is similar.
$\qed$

In view of the first part of the paper we can state that any pure 
entangled state and any bound entangled UPB state can be detected
with the methods of Lemma 2. 
Identifying more general classes of states which
allow a detection leads to the difficult problem 
of characterizing the possible 
$\sum_k p_k \gamma(\varrho^A_k \otimes \varrho^B_k, M_i)$
in Eq. (\ref{l2a}), resp. the $\kappa_{A/B}$ in 
Eqns. (\ref{p8a}, \ref{p8b}).
We only know some properties of them:
Taking $A_i = B_i =\sigma_i, i=x,y,z$
for two qubits, we know that  
$\sum_i \delta^2(A_i)\geq 2$ \cite{hofmann1}, 
thus $Tr(\kappa_A)\geq 2.$
Applying this to the Werner states 
$\varrho(p)=p\ketbra{\psi}+(1-p)\eins/4$ with 
$\ket{\psi}= (\ket{01}-\ket{10})/\sqrt{2}$
one can calculate that they are detected 
via Eq. (\ref{p8a})
for $p\geq 1/\sqrt{3}.$

In conclusion, we have developed a family of necessary separability 
criteria based on inequalities for variances. We have shown that they
are strong enough to detect the entanglement in various experimental 
relevant situations. We have formulated an equivalent criterion in 
terms of covariance matrices, which enabled us to connect 
continuous variable systems with finite-dimensional systems.
The question whether there are entangled states which
do not violate any uncertainty relation, is a very interesting 
one; here, it remains open. 

We wish to thank D.~Bruß, G.~Giedke, H.~Hofmann, F.~Hulpke, 
P.~Hyllus, M.~Lewenstein,
U.~Poulsen, A.~Sanpera, and G.~T\'oth for discussions. 
This work has been supported 
by the DFG.


\begin{thebibliography}{99}

\bibitem{criteria} 
A.C. Doherty, P.A. Parrilo, and  F.M. Spedalieri, 
                   Phys. Rev. Lett. {\bf 88}, 187904 (2002);
                   O. Rudolph, quant-ph/0202121. K. Chen and L. Wu, 
                   Quant. Inf. Comp.  {\bf 3}, 193 (2003);
                   M. Horodecki, P. Horodecki, and 
                   R. Horodecki, quant-ph/0206008;
                   for a review see
                   D. Bru\ss~{\it et al.}, J. Mod. Opt. {\bf 49}, 1399 (2002).

\bibitem{bell} See, e.g. A. Peres, Found. Phys. {\bf 29}, 589 (1999).

\bibitem{witnesses} M. Horodecki, P. Horodecki, and  R. Horodecki, 
                    Phys. Lett. A {\bf 223}, 1 (1996);
                    B.M. Terhal, {\it ibid.} {\bf 271}, 319 (2000);
                    M. Lewenstein {\it et al.}, Phys. Rev. A {\bf 62}, 
                    052310 (2000).

\bibitem{alle} M.D. Reid and P.D. Drummond, Phys. Rev. Lett. {\bf 60}, 
               2731 (1988); M.D. Reid, Phys. Rev A {\bf 40}, 913 (1989);
               L.-M. Duan {\it et al.}, Phys. Rev. Lett. {\bf 84}, 2722 (2000);
               S. Mancini {\it et al.}, {\it ibid.} {\bf 88}, 120401 (2002);
               N. Korolkova {\it et al.}, Phys. Rev. A {\bf 65}, 052306 (2002);
               V. Giovannetti {\it et al.}, {\it ibid.} {\bf 67}, 022320 (2003); 
               P. van Loock and A. Furusawa, {\it ibid.} 
               {\bf 67}, 052315 (2003); G. T\'oth, C. Simon, and  J.I. Cirac, 
               {\it ibid.} {\bf 68}, 062310 (2003).

\bibitem{hofmann1} H.F. Hofmann and S. Takeuchi, Phys. Rev. A {\bf 68}, 
                   032103 (2003).

\bibitem{hofmann2} H.F. Hofmann, Phys. Rev. A {\bf 68}, 034307 (2003)

\bibitem{werner} R.F. Werner and  M.M. Wolf, Phys. Rev. Lett. {\bf 86}, 3658
                 (2001).

\bibitem{giedke1} G. Giedke {\it et al.}, Phys. Rev. Lett. {\bf 87}, 167904
                  (2001).

\bibitem{galindo} A. Galindo and P. Pascual: {\it Problemas de mec\'anica
                  cu\'antica}, (Eudema Universidad Manuales, Madrid
                  1989), p. 56. 

\bibitem{exdet} B.M. Terhal, Theoret. Comput. Sci. {\bf 287}, 313 (2002).
                O. Gühne {\it et al.}, Phys. Rev. A {\bf 66}, 062305 (2002).

\bibitem{fdm} M. Barbieri {\it et al.}, Phys. Rev. Lett. {\bf 91}, 227901 (2003); M. Bourennane 
              {\it et al.}, quant-ph/0309043.

\bibitem{upb1} C.H. Bennett {\it et al.}, Phys. Rev. Lett. {\bf 82}, 5385 
              (1999).

\bibitem{duer1} W. Dür, G. Vidal, and J.I. Cirac, Phys. Rev. A, {\bf 62}, 
                062314 (2000).

\bibitem{acin} A. Ac{\'\i}n {\it et al.}, Phys. Rev. Lett. {\bf 87}, 040401 
              (2001).

\bibitem{ustron} O. Gühne and P. Hyllus, Int. J. Theor. Phys. {\bf 42}, 1001 
                 (2003). 

\bibitem{gernot} See, e.g. I. Montvay and G. Münster: {\it Quantum Fields 
                 on a Lattice}, (Cambridge University Press, 1994).

\bibitem{wilcox} R.M. Wilcox, J. Math. Phys. {\bf 8}, 962 (1967). 

\bibitem{scharlau} See, e.g. F. Hirzebruch and  W. Scharlau, 
                  {\it Einführung in die Funktionalanalysis}, 
                  (BI, Mannheim, 1971).

\bibitem{gezaun} Similar witness-like constructions have been considered 
                 before for CMs in continuous variables. 
                 (G. Giedke and M. Lewenstein,  
                 private communication, 2002).

\bibitem{oplus} The symbol $A \oplus B$ denotes a $2 \times 2$-block matrix 
consisting in $A$ and $B$ on the diagonal and zero matrices elsewhere.  

\bibitem{pseudoinverse} Here, $X^{-1}$ denotes the pseudo-inverse of 
the matrix $X,$ {\it i.e.} the inversion on the range. 

\end{thebibliography}
\end{document}